\documentstyle[12pt]{article}
\begin{document}

\begin{center}

{\LARGE \bf Neutral perfect fluids of Majumdar-type \\
\vspace{0.5cm}in general relativity}

\vskip 1truecm

{\bf Victor Varela}

Centro de F\'{\i}sica Te\'{o}rica y Computacional,\\
Escuela de F\'{\i}sica, Facultad de Ciencias,\\
Universidad Central de Venezuela,\\
AP 47270, Caracas 1041-A, Venezuela\\
{\em E-mail: {\tt vvarela@fisica.ciens.ucv.ve}}\\

\end{center}

\vskip 1truecm

\abstract{We consider the extension of the Majumdar-type class of
static solutions for the Einstein-Maxwell equations, proposed by
Ida to include charged perfect fluid sources. We impose the
equation of state $\rho+3p=0$ and discuss spherically symmetric
solutions for the linear potential equation satisfied by the
metric. In this particular case the fluid charge density vanishes
and we locate the arising neutral perfect fluid in the
intermediate region defined by two thin shells with respective
charges $Q$ and $-Q$. With its innermost flat and external
(Schwarzschild) asymptotically flat spacetime regions, the resultant
condenser-like geometries resemble solutions discussed by Cohen
and Cohen in a different context. We explore this relationship and
point out an exotic gravitational property of our neutral perfect
fluid. We mention possible continuations of this study to embrace
non-spherically symmetric situations and higher dimensional
spacetimes.}

\vskip 1truecm

Our previous contribution \cite{vv1} was devoted to the study of
static charged dust solutions of the Einstein-Maxwell equations.
The derivation of the solutions was based on three main
assumptions: (i) the spacetime metric has the conformastatic form
\begin{equation}
ds^{2}=-V^2\;dt^{2}+\frac{1}{V^2}\;h_{ij}dx^{i}dx^{j},
\label{synge}
\end{equation}
where Latin indices indicate $1, 2 ,3$; (ii) the background
three-dimensional metric $h_{ij}$ is Euclidean; (iii) $h_{ij}$ and
$V$ depend only on the space-like coordinates $x^{1}, x^{2},
x^{3}$. As a consequence, the corresponding static solutions of
the Einstein-Maxwell-charged dust equations satisfied a linear
relation between $V$ and the Coulombian potential $A_{0}$. We will
refer to solutions of the field equations with such a linear
relationship as being of Majumdar-type. See Guilfoyles' paper
\cite{guilfoyle} for a discussion of these geometries within the
framework of the Weyl class of solutions \cite{weyl}. More recent
papers by Vogt and Letelier \cite{vogt}, Kleber {\it et al.}
\cite{kleber}, Ivanov \cite{ivanov}, and Wickramasuriya
\cite{wick} deal with interesting physical and mathematical
aspects of Majumdar-type solutions.

When the static charged dust source is Majumdar-type, the field
equations imply that the charge density $\sigma$ and the energy
density $\rho$ satisfy the relation $\sigma = \pm \rho$ and the
metric function $\lambda = \frac{1}{V}$ is a solution of the
potential equation
\begin{equation}
\nabla^{2}_{(h)}\lambda+4\pi\rho\lambda^{3}=0,
\label{potequ}
\end{equation}
where $\nabla^{2}_{(h)}$ denotes the Laplacian operator associated
to the background space with flat metric $h_{ij}$. G\"{u}rses
\cite{gur} discussed internal solutions without space-like Killing
vectors when Eq. (\ref{potequ}) is linear.

The extension of the Majumdar-type class of solutions to include
static charged perfect fluid sources is easily achieved using a
modified version of the charged dust procedure outlined in
\cite{vv1}. In this case the conformastatic metric (\ref{synge})
is assumed again, but $h_{ij}$ is allowed to have non-vanishing
Riemannian curvature.

In the static perfect fluid case with metric (\ref{synge}) and
four-velocity $u^{\mu}=\frac{1}{V}\delta^{\mu}_{0}$ (we use Greek
indices for the range $0, 1, 2, 3$), the components of the matter
energy-momentum tensor
\begin{equation}
M_{\mu\nu}=\rho u_{\mu}u_{\nu}+ p \left( g_{\mu\nu}+u_{\mu}u_{\nu} \right)
\label{pftmunu}
\end{equation}
are given by
\begin{eqnarray}
M_{00}=\rho V^{2},&
M_{0i}=0,&
M_{ij}=p \frac{h_{ij}}{V^{2}}.
\label{Mcomps}
\end{eqnarray}

As in the charged dust case, the electrostatic forms of the
four-potential $A_{\mu}$ and the four-current $J^{\mu}$ are given
by
\begin{equation}
A_{\mu}=A_{0}(x^{i})\delta^{0}_{\mu},
\label{Astatic}
\end{equation}
\begin{equation}
J^{\mu}=\frac{\sigma (x^{i})}{V}\delta^{\mu}_{0}.
\label{Jstatic}
\end{equation}
The definitions of the electromagnetic tensor
\begin{equation}
F_{\mu\nu}=\partial_{\mu}A_{\nu}-\partial_{\nu}A_{\mu}
\label{maxdef}
\end{equation}
and the Maxwell energy-momentum tensor
\begin{equation}
E_{\mu\nu}=\frac{1}{4\pi}\left(F_{\mu\alpha}F_{\nu}^{ \alpha}-
\frac{1}{4}g_{\mu\nu}F_{\alpha\beta}F^{\alpha\beta}\right)
\label{maxtmunu}
\end{equation}
lead to the components
\begin{equation}
E_{00}=\frac{1}{8\pi}V^{2}h^{ij}\partial_{i}A_{0} \partial_{j}A_{0},
\label{E00}
\end{equation}
\begin{equation}
E_{0i}=0,
\label{E0i}
\end{equation}
\begin{equation}
E_{ij}=\frac{1}{4\pi}\left(-\frac{1}{V^{2}}\partial_{i}A_{0}\partial_{j}A_{0}+
\frac{1}{2}\frac{1}{V^{2}}h_{ij}h^{kl}\partial_{k}A_{0}
\partial_{l}A_{0}\right),
\label{Eij}
\end{equation}
where $h^{ij}$ is the inverse of $h_{ij}$.

Using Eq. (\ref{synge}) we determine the components of the
Einstein tensor, given by
\begin{equation}
G_{00}=-3V^{2}h^{ij}\partial_{i}V\partial_{j}V+2V^{3}\nabla^{2}_{(h)}V+
\frac{1}{2}V^{4}R_{(h)},
\label{G00}
\end{equation}
\begin{equation}
G_{0i}=0,
\label{G0i}
\end{equation}
\begin{equation}
G_{ij}=R_{(h)ij}-\frac{2}{V^{2}}\partial_{i}V\partial_{j}V+
h_{ij}\left(\frac{1}{V^{2}}h^{mk}\partial_{m}V\partial_{k}V-\frac{1}{2}R_{(h)}\right),
\label{Gij}
\end{equation}
where $R_{(h)}$ and $R_{(h)ij}$ respectively denote the Ricci
scalar and the Ricci tensor constructed with $h_{ij}$.

When we combine the above results with the Einstein equations
\begin{equation}
G_{\mu\nu}=8\pi \left( M_{\mu\nu}+E_{\mu\nu} \right)
\label{eins}
\end{equation}
the following non-trivial equations are obtained:
\begin{equation}
\frac{1}{2} V^{4} R_{(h)}-3V^{2}h^{ij}\partial_{i}V\partial_{j}V+2V^{3}\nabla^{2}_{(h)}V=
V^{2}h^{ij}\partial_{i}A_{0}\partial_{j}A_{0}+8\pi\rho V^{2},
\label{eins00}
\end{equation}
\begin{equation}
V^{2} G_{(h)ij} -2\partial_{i}V\partial_{j}V+h_{ij}h^{kl}\partial_{k}V\partial_{l}V=
-2\partial_{i}A_{0}\partial_{j}A_{0}+h_{ij}h^{kl}\partial_{k}A_{0}\partial_{l}A_{0}+8\pi p h_{ij},
\label{einsij}
\end{equation}
where $G_{(h)ij}$ is the Einstein tensor constructed with
$h_{ij}$.

A great simplification of these equations is achieved if we assume
(i) that the three-dimensional background space with metric
$h_{ij}$ is maximally symmetric, and (ii) that $h_{ij}$, $p$, and
$V$ are related by
\begin{equation}
G_{(h)ij}=8 \pi p \frac{h_{ij}}{V^{2}}.
\label{eins3d}
\end{equation}
As a consequence, the extended Majumdar-type solutions satisfy
\begin{equation}
R_{(h)}= constant,
\label{rhconst}
\end{equation}
and the pressure is given by
\begin{equation}
p=-\frac{R_{(h)}V^2}{48 \pi }.
\label{presure}
\end{equation}
Combining Eqs. (\ref{einsij}) and (\ref{eins3d}), and using
$h^{i}_{i}=3$ we obtain
\begin{equation}
h^{kl}\partial_{k}V\partial_{l}V=h^{kl}\partial_{k}A_{0}\partial_{l}A_{0}.
\label{cuads}
\end{equation}
Plugging this result and Eq. (\ref{eins3d}) into Eq.(\ref{einsij})
we find that
\begin{equation}
\partial_{i}V\partial_{j}V=\partial_{i}A_{0}\partial_{j}A_{0},
\label{ijred}
\end{equation}
which integrates up to
\begin{eqnarray}
V=\kappa A_{0}+C,&
\kappa^{2}=1,
\label{linear}
\end{eqnarray}
for some constant $C$. Using Eqs. (\ref{eins00}), (\ref{presure}),
(\ref{linear}), and the identity
\begin{equation}
\nabla^{2}_{(h)}V=\frac{2}{V}h^{ij}\partial_{i}V\partial_{j}V
-V^{2}\nabla^{2}_{(h)}\left(\frac{1}{V}\right)
\label{identi}
\end{equation}
we get the extended potential equation
\begin{equation}
\nabla^{2}_{(h)}\lambda+4\pi\left(\rho+3 p \right)\lambda^{3}=0.
\label{potential}
\end{equation}

We can use Eq. (\ref{linear}) to eliminate $A_{0}$ in the
non-trivial Maxwell equation
\begin{equation}
\frac{1}{\sqrt{h}}\;\partial_{j}\left(\sqrt{h}\; h^{jk}\;
\frac{\partial_{k}A_{0}}
{V^{2}}\right)=\frac{4\pi \sigma}{V^3},
\label{nontriv}
\end{equation}
where $h$ is the determinant of $h_{ij}$. The resultant expression
is
\begin{equation}
\nabla^{2}_{(h)}\lambda + \frac{4\pi\sigma}{\kappa}\lambda^{3}=0.
\label{potmax}
\end{equation}
Comparing Eqs. (\ref{potential}) and (\ref{potmax}) we conclude
that $\sigma$, $\rho$, and $p$ are related by
\begin{equation}
\sigma=\kappa \left(\rho+3 p \right).
\label{idarel}
\end{equation}

Equation (\ref{idarel}) characterizes the class of static charged
perfect fluid solutions of Majumdar-type. It was obtained by Ida
\cite{ida} using a different formalism. Guilfoyle \cite{guilfoyle}
studied properties of this class of solutions. Cho {\it et al.}
\cite{cho} and Yazadjiev \cite{yaza} have generalized Eqs.
(\ref{potential}) and (\ref{idarel}) with the inclusion of a
dilaton field.

We have seen that extended Majumdar-type solutions must satisfy
Eq. (\ref{presure}). Hence non-zero values for constant $R_{h}$
cause metric singularities at spacetime points where $p=0$. This
fact complicates the construction of bounded, perfect fluid
sources surrounded by ordinary vacuum, characterized by
$\rho=p=0$. However we easily conceive composite sources where
these geometries describe only internal regions, so that metric
singularities can be avoided. In this work we shall consider
composite models that include Majumdar-type perfect fluids in the
case of spherical symmetry. (Composite charged dust spheres have
been considered in references \cite{vv1} and \cite{kleber}.)

Equations (\ref{potential}) and (\ref{idarel}) relate perfect
fluid properties, charge distribution and analytic structure of
the solutions. For example, if $\rho + p = 0$ then $M_{\mu\nu}$ is
isotropic and $\lambda$ is an eigenfunction of $\nabla^{2}_{(h)}$
with eigenvalue $\frac{R_{h}}{6}$. We are particularly interested
in perfect fluids with equation of state
\begin{equation}
\rho+3 p=0.
\label{eos}
\end{equation}
Majumdar-type sources with this property are necessarily neutral
as a consequence of Eq. (\ref{idarel}). In this case $\lambda$ is
a solution of the Laplace equation in a background space of
constant curvature.

Using spherical coordinates Eq. (\ref{synge}) can be written as
\begin{equation}
ds^{2}=-V^{2}dt^{2}+\frac{1}{V^{2}} \frac{1}{\left(1+\frac{K
r^{2}}{4} \right)^{2}} \left(dr^2+r^{2} d\theta^{2} + r^{2}
\sin\theta^{2} d\phi^2 \right),
\label{dsce}
\end{equation}
where $K$ is a constant. Also Eq. (\ref{presure}) takes the form
\begin{equation}
p=-\frac{K V^2}{8 \pi },
\label{presion}
\end{equation}
so that $\rho$ and $K$ always have the same sign.

The dominant energy condition implies that
\begin{eqnarray}
\hat{\rho} \ge 0,&
-\hat{\rho} \le \hat{p}_{i} \le \hat{\rho},
\label{decorto}
\end{eqnarray}
where $\hat{\rho}$ and $\hat{p}_{i}$, $i=1,2,3$, are the
components of the covariant, total energy-momentum tensor in an
orthonormal base. The spherically symmetric expressions for these
components have the form
\begin{equation}
\hat{\rho}=\frac{1}{8 \pi}\left(1+\frac{K r^{2}}{4} \right)^{2}
\left(\frac{\partial A_{0}}{\partial r}\right)^2 + \rho,
\label{hat00}
\end{equation}
\begin{equation}
\hat{p}_{1}=-\frac{1}{8\pi}\left(1+\frac{K r^{2}}{4} \right)^{2}
\left(\frac{\partial A_{0}}{\partial r}\right)^2 + p,
\label{hat11}
\end{equation}
\begin{equation}
\hat{p}_{2}=\frac{1}{8\pi}\left(1+\frac{K r^{2}}{4} \right)^{2}
\left(\frac{\partial A_{0}}{\partial r}\right)^2+p,
\label{hat22}
\end{equation}
\begin{equation}
\hat{p}_{3}=\frac{1}{8 \pi}\left(1+\frac{K r^{2}}{4} \right)^{2}
\left(\frac{\partial A_{0}}{\partial r}\right)^2 + p.
\label{hat33}
\end{equation}

Using Eqs. (\ref{decorto})-(\ref{hat33}) we conclude that the
dominant energy condition is equivalent to
\begin{eqnarray}
\rho \ge 0,&
-\rho \le {p} \le \rho.
\label{decmatter}
\end{eqnarray}
We observe that every spherically symmetric Majumdar-type solution
describing matter with equation of state (\ref{eos}) will be
compatible with Eq. (\ref{decmatter}) whenever $p < 0$, or
equivalently $K > 0$.

In the spherically symmetric case $\lambda= \lambda (r)$, and
the general solution of the Laplace equation
\begin{equation}
\nabla^{2}_{(h)}\lambda =0
\label{lapeq}
\end{equation}
is given by
\begin{equation}
\lambda = a+b\left(\frac{1}{r}-\frac{Kr}{4}\right),
\label{lapsol}
\end{equation}
where $a$ and $b$ are real integration constants.

Projecting the electromagnetic tensor onto the unit vector
$n=\left(1+\frac{Kr^{2}}{4} \right) V \frac{\partial}{\partial r}$
and the four-velocity $u=\frac{1}{V}\frac{\partial}{\partial t}$,
and using the linear relationship between $A_{0}$ and $V$ we
obtain the scalar
\begin{equation}
E = F_{\mu \nu}n^{\mu}u^{\nu}=\kappa \frac{\partial V}{\partial
r}\left(1+\frac{Kr^{2}}{4} \right).
\label{defE}
\end{equation}
Additionally, we use Eq. (\ref{lapsol}) to derive the expression
\begin{equation}
E = \kappa b
\frac{V^{2}}{r^{2}}\left(1+\frac{Kr^{2}}{4}\right)^{2}.
\label{evalE}
\end{equation}

The invariant area of the 2-sphere $t=constant, r=constant$ with
unit normal $n$ is given by
\begin{equation}
S=\frac{4\pi r^{2}}{V^{2}\left(1+\frac{Kr^{2}}{4}\right)^{2}}.
\label{S}
\end{equation}
Combining Eqs. (\ref{evalE}) and (\ref{S}) we obtain
\begin{equation}
ES=4\pi \kappa b.
\label{ES}
\end{equation}
In view of the Gauss theorem \cite{wald}, we derive the expression
\begin{equation}
b=\kappa Q,
\label{b}
\end{equation}
where $Q$ is the total charge enclosed by the 2-sphere.

We devote the rest of this work to the construction of bounded
sources for asymptotically flat geometries using the neutral
perfect fluid solution (\ref{lapsol}). We restrict our attention
to $C^{0}$ junctions with both the internal and external metrics
represented in isotropic coordinates.

First we consider the $C^{0}$ junction of the internal solution
given by Eqs. (\ref{dsce}) and (\ref{lapsol}) to the Schwarzschild
solution
\begin{equation}
ds^{2}= -\frac{\left(1-\frac{M}{2r}\right)^{2}}
{\left(1+\frac{M}{2r}\right)^{2}}dt^{2} +
\left(1+\frac{M}{2r}\right)^{4}\left(dr^2+r^{2} d\theta^{2} +
r^{2} \sin\theta^{2} d\phi^2 \right)
\label{schw}
\end{equation}
at $r = r_{2} > 0 $. The continuity of the metric coefficients and
Eq. (\ref{b}) imply expressions
\begin{equation}
K=\frac{M^{2}}{r_{2}^{4} \left(1-\frac{M^{2}}{4r_{2}^{2}}\right)}
\label{Kformula}
\end{equation}
and
\begin{equation}
a=\frac{\left(1+\frac{M}{2r_{2}}\right)^{2}-\frac{\kappa
Q}{r_{2}}\left(1-\frac{M^{2}}{2r_{2}^{2}}\right)}{1-\frac{M^{2}}{4r_{2}^{2}}}.
\label{aformula}
\end{equation}
In view of Eq. (\ref{Kformula}), $M<2r_{2}\Rightarrow K>0$. In
this case Eq. (\ref{presion}) implies $p<0$, and $\rho$ is
positive as a consequence of Eq. (\ref{eos}). We conclude that the
dominant energy condition is satisfied by the neutral perfect
fluid whenever the Schwarzschild horizon is excluded from the
external solution.

The external geometry is determined by the Schwarzschild metric
(\ref{schw}), and we expect the hyper-surface $r=r_{2}$ to define
a thin shell with charge $-Q$, so that the composite source has
zero net charge. Indeed, the singular character of this
hyper-surface shows up when we calculate the integrated stresses
\begin{equation}
\Sigma^{\alpha}_{\beta}= \lim_{\epsilon \to
0}\int\limits_{r_{2}-\epsilon}^{r_{2}+\epsilon}
T^{\alpha}_{\beta}d\hat{r},
\label{intestress}
\end{equation}
where $T^{\alpha}_{\beta}$ and $d\hat{r}$ denote the components of
the total (mixed) energy-momentum tensor and the infinitesimal,
invariant (spatial) distance, respectively. In order to evaluate
these integrals we use the Einstein equations, assume the general
form of the invariant element in isotropic coordinates
\begin{equation}
ds^{2}=-e^{-2\gamma (r)}dt^{2}+e^{2\alpha (r)}\left(dr^2+r^{2}
d\theta^{2} + r^{2} \sin\theta^{2} d\phi^2 \right),
\label{isometric}
\end{equation}
use the expression $d\hat{r}=e^{\alpha (r)}dr$, and note that only
terms in the mixed Einstein tensor with second derivatives of
$\gamma (r)$ and $\alpha (r)$ can contribute \cite{problem}. Using
the internal metric (\ref{dsce}) combined with Eqs.
(\ref{lapsol}), (\ref{b}), (\ref{Kformula}), (\ref{aformula}) for
$r < r_{2}$, the external metric (\ref{schw}) for $r
> r_{2}$, and the corresponding formulae derived in \cite{problem}
we finally obtain
\begin{equation}
\Sigma^{0}_{0}=\frac{8\kappa r_{2}^{3} Q + 4 M^{3} r_2 + M^{4} -
8r_{2}^{3} M}{2\pi r_{2}
\left(4r_{2}^{2}+4r_{2}M+M^{2}\right)^{2}},
\label{s00}
\end{equation}
\begin{equation}
\Sigma^{1}_{1}=0,
\label{s11}
\end{equation}
\begin{equation}
\Sigma^{2}_{2}=\Sigma^{3}_{3}=
\frac{\left(8r_{2}^{2}-M^{2}\right)M^{2}}{4\pi r_{2}
\left(4r_{2}^2+4r_{2}M+M^{2}\right)\left(4r_{2}^{2}-M^{2}\right)}.
\label{s2233}
\end{equation}

To complete the construction of the bounded source with $C^{0}$
metric and zero net charge, we impose the junction of the internal
solution (\ref{lapsol}) to the flat metric
\begin{equation}
ds^{2}=-dt^2 + dr^2 + r^{2} \left( d\theta^{2} + \sin\theta^{2}
d\phi^2 \right)
\label{flatmet}
\end{equation}
at $r=r_{1}$, with $0 < r_{1} < r_{2} $. We expect the
hyper-surface $r=r_{1}$ to define a thin shell with charge $Q$. In
fact, if we evaluate the integrated stresses (\ref{intestress}) at
$r=r_{1}$ then we obtain more complicated expressions that impose
finite values for $\Sigma^{0}_{0}$, $\Sigma^{2}_{2}$, and
$\Sigma^{3}_{3}$ on this hyper-surface.

We summarize the construction details of the source as follows.

The source is composed by (i) the innermost flat region, which is
empty in the ordinary sense ($\rho=p=0$) and is defined for $0
\leq r < r_{1}$; (ii) the intermediate region $r_{1} < r < r_{2}$,
which contains a neutral perfect fluid with equation of state
$\rho + 3p=0$ and $\rho>0$; (iii) the external region $r
> r_{2}$, with $\rho=p=0$ and asymptotically flat (Schwarzschild)
metric. Hyper-surfaces $r=r_{1}$ and $r=r_{2}$ constitute thin
shells with charges $Q$ and $-Q$, respectively. The Maxwell tensor
vanishes only in the innermost and external regions.

Our composite source resembles the self-gravitating electrical
condenser presented by Cohen and Cohen \cite{cohencohen}. Two
important differences are (i) the use of curvature coordinates by
these authors; (ii) the imposition of ordinary vacuum ($\rho=p=0$)
in the intermediate region of their condenser, where only the
electric field contributes to the total energy-momentum tensor.

Combining Eqs. (\ref{lapsol}), (\ref{S}), (\ref{b}),
(\ref{Kformula}), (\ref{aformula}), and defining
$\mu=\frac{M}{2r_{2}}$, $\beta=\frac{Q}{r_{2}}$,
$\alpha=\frac{r_{2}}{r_{1}}$, we derive the following expression
for the invariant area
\begin{equation}
\tilde{S}=\frac{\left[\frac{\left(1+\mu^{2}\right)-\kappa \beta
\left(1-2\mu^{2}\right)}{1-\mu^{2}}x+\kappa \alpha
\beta\left(1-\frac{\mu^{2}x^{2}}
{\alpha^{2}\left(1-\mu^{2}\right)}\right)\right]^{2}}
{\left(1+\frac{\mu^{2}x^{2}}{\alpha^{2}\left(1-\mu^{2}\right)}
\right)^{2}}, \label{Stilde}
\end{equation}
where $\tilde{S}=\frac{S}{4 \pi r_{1}^{2}}$ and
$x=\frac{r}{r_{1}}$. Certainly, $\tilde{S}$ has a complicated
dependence on parameters $\alpha$, $\beta$, $\kappa$ and $\mu$.

Distinct behaviors of $\tilde{S}$ arise when we examine few simple
examples. If we choose $\alpha=2$, $\beta=1$, $\kappa=1$ and
$\mu=0.1, 0.2, 0.3, 0.4, 0.5$, then we obtain monotonously
increasing $\tilde{S}$ functions for $x\in [1,2]$. Nevertheless
$\tilde{S}$ has local maxima when $\mu=0.6, 0.7, 0.8$, and the
choice $\mu=0.9$ leads to a monotonously decreasing $\tilde{S}$ in
the same interval. If $\kappa=-1$ and the values of $\alpha$ and
$\beta$ remain the same, then we obtain monotonously increasing
$\tilde{S}$ functions for $x\in [1,2]$ corresponding to the
choices $\mu=0.1, 0.2, 0.3, 0.4, 0.5, 0.6, 0.7, 0.8$. In this case
$\tilde{S}$ has a local maximum when $\mu=0.9$. On the other hand,
the model defined by $\alpha=10$, $\beta=1$, $\kappa=-1$,
$\mu=0.5$ develops a singularity at $x \approx 2.663$.

If we represent the metric of the intermediate region in curvature
coordinates $t,R,\theta,\phi$, then we conclude that $S=4\pi
R^{2}$. Therefore solutions with monotonously increasing invariant
area in the interval $x\in [1,\alpha]$ have the simplest
interpretation in terms of the radial marker $R$, and are closely
related to the Cohen and Cohen solution.

We have filled the intermediate region of our composite source
with an electric field and a neutral perfect fluid. In contrast,
the intermediate region of the Cohen and Cohen condenser contains
only an electric field. Surprisingly, this difference diminishes
when we have a closer look at the gravitational properties of
perfect fluids with equation of state $\rho + 3p=0$.

As discussed by Ponce de Le\'{o}n \cite{pdl1,pdl2}, the use of
curvature coordinates entails an expression for the gravitational
mass $M_{G}(R)$ inside a sphere of radius $R$ (the Tolman-Wittaker
formula). (Gr{\o}n \cite{gron} considered the discontinuities in
$M_{G}(R)$ caused by thin shell singularities.) The crucial point
is that $\rho+3p$ determines the contribution of the matter to
$M_{G}(R)$ in the case of isotropic pressures. The analysis of
these authors suggest that our neutral perfect fluid has no direct
effect on local gravitational interactions. (Matter with the odd
equation of state $\rho+3p=0$ was considered by Gott and Rees in
the context of cosmic strings \cite{gottrees}. Wesson related
$\rho + 3p=0$ to the existence of zero-point fields \cite{wesson}.
Dadhich and Narayan \cite{dad} regarded various topological
defects in connection with this equation of state.)

Equations (\ref{presion}) and (\ref{Kformula}) show the effect of
mass $M$ on pressure $p$, and we presume that our fluid affects
the contribution of the charged thin shells to the gravitational
mass of the system (via boundary conditions), in analogy with the
induction of surface charge on the plates of a condenser by a
neutral, polarized dielectric substance. Systematic use of
curvature coordinates should lead to an expression for $M$ as a
function of $Q$ and the condenser's capacity, modified by the
exotic matter presence in the intermediate region (see Eq. (30) in
\cite{cohencohen}).

We finally remark that both the neutral fluid condition and the
simplification that leads to the Laplace equation for $\lambda$
are consequences of $\rho+3p=0$ even in the absence of space-like
Killing vectors. Notably, Eqs. (21) and (24) in \cite{cho} suggest
that the equation of state $\rho + \frac{N}{N-2} p = 0 $ plays an
analogous role in $N+1$ dimensional spacetimes.

\section*{Acknowledgements}
The author thanks his colleagues P. J. Arias, A. Bellor\'{\i}n,
and L. Leal for timely computational support.

\end{document}